**Impact of size mismatch induced quenched disorder on phase fluctuation and low field magnetotransport in polycrystalline $Nd_{0.58-x}Gd_xSr_{0.42}MnO_3$**

**Manoj K. Srivastava[1], Ravikant Prasad[1], P. K. Siwach[1], M. P. Singh[2] and H. K. Singh[1#]**

[1]*National Physical Laboratory (CSIR), Dr. K. S. Krishnan Marg, New Delhi-110012, India*
[2]*Department of Materials Science and Engineering, University of Washington, Seattle, Washington 98195-2120, USA*

Abstract

We report the magnetic and electrical transport properties of the bulk polycrystalline $Nd_{0.58-x}Gd_xSr_{0.42}MnO_3$ (x ~ 0.0, 0.04, 0.08, 0.12, 0.16, 0.20, 0.25, 0.30, 0.35 and 0.42) synthesized by solid state reaction at 1350 °C. All the samples are single phase and have grain size ~1-2 micrometer. The variation of magnetic and magnetotransport properties as a function of the variance $\sigma^2$ show some interesting trends. As $\sigma^2$ increases, the paramagnetic-ferromagnetic (PM-FM) transition shows a gradual decrease and broadening, while the decrease in insulator metal transition (IMT) is accompanied by sharpening. However, near equality of $T_C$ and $T_{IM}$ at intermediate values of $\sigma^2$ could be understood in terms of the competing quenched and the grain boundary disorder. The variation of peak MR (maximum MR around $T_C/T_{IM}$) with $\sigma^2$ shows that maximum low field MR≈35 % at H=3 kOe (68 % at H=10 kOe) is centered around $\sigma^2$ = 0.009857 $Å^2$ (x=0.25). This shows that huge intrinsic MR can be obtained at relatively higher temperatures and lower magnetic fields in the region of enhanced phase fluctuations.

[#] **Electronic mail: hks65@nplindia.org**

## Introduction

Phase coexistence due the occurrence of several magneto-electronic phases, such as paramagnetic insulator (PMI), ferromagnetic metal (FMM), ferromagnetic insulator (FMI), antiferromagnetic metal/insulator (AFMM/AFMI) and charge/orbital order insulator (COI) has been recognized as the intrinsic property of the doped rare earth manganites, represented by general formula $RE_{1-x}AE_xMnO_3$ (RE=La, Pr, Nd etc., and AE=Ca, Ba, Sr, etc.).[1-5] These compounds also exhibit colossal magnetoresistance (CMR), which is now believed to be related intimately to COI state.[1-5] Therefore the strong magneto-electric phase coexistence and the keen competition between the COI and the FMM phases are now crucial elements in the potential explanation of CMR.[2-4] However, the magneto-electrical properties, particularly the phase coexistence/fluctuation cannot be determined solely by the 3d electron transfer interaction or one-electron bandwidth (W).[2] The structural and magneto-electrical properties can be different even for the identically distorted perovskite lattice structure, the former of which is generally determined by the tolerance factor or the A-site averaged ionic radius alone. Attfield and Rodriguez-Martinez[6-8] conclusively demonstrated that the FM transition temperature ($T_C$) as well as the metal–insulator transition temperature ($T_{IM}$) depends strongly not only on the average ionic radius but also on the size mismatch of the trivalent rare-earth and divalent alkaline-earth ions of the A-site of the $ABO_3$ type structure. The size mismatch representing the local lattice distortion transmitted from the randomly substituted A-site is measured by the variance, $\sigma^2 = \Sigma y_i r_i^2 - <r_A>^2$, where $y_i$ is the fraction of the $i^{th}$ cation, $r_i$ is its radius and $<r_A>$ is the average radius of the cations at the rare earth site. The local distortion arising from the difference in ionic radii, that is, $\sigma^2$ and/or the random Coulomb potential due to the trivalent/divalent ion mixture is the source of the quenched disorder. Here, quenched disorder means the temperature-independent atomic-scale local inhomogeneity producing randomness in potential energy, transfer energy, electron number, exchange interaction, etc. In the hole-doped perovskite manganites, therefore, this kind of quenched disorder or the potential randomness is unavoidable except for some special cases.[2]

Size and hence quenched disorder has been shown to induce variety of modifications in the ground state of the manganites.[2] However, the nature of the modification depends on the carrier concentration, i.e., the concentration of the divalent cation. Thus size disorder introduces variety of phenomena, such as, metamagnetism and ferromagnetic-metal to cluster glass-insulator transition[9-10] at x≈0.5, collapse of the CO-OO state at x≈0.45[11] and the Griffiths phase[12] in low and intermediate W manganites. The magneto-electric phase evolution as a function of size disorder ($\sigma^2$) is expected to strongly influence the low field magnetotransport properties. In the present work we report the effect of size mismatch induced disorder on phase fluctuation and low field magneto electrical properties of intermediate W manganite $Nd_{0.58-x}Gd_xSr_{0.42}MnO_3$. The size disorder has been introduced by partial substitution the larger $Nd^{3+}$ (r=1.27 Å) by smaller $Gd^{3+}$ (r=1.21 Å) cation. The $<r_A>$ and $\sigma^2$ were calculated using twelve-coordinate ionic radii given by Shannon.[13]

**Experimental**

Polycrystalline $Nd_{0.58-x}Gd_xSr_{0.42}MnO_3$ (x ~ 0.0, 0.04, 0.08, 0.12, 0.16, 0.20, 0.25, 0.30, 0.35 and 0.42) samples were synthesized by the conventional solid state reaction method. The desired amounts of high purity (99.99) $Nd_2O_3$, $Gd_2O_3$, $SrCO_3$ and $MnCO_3$ were weighed and mixed homogeneously. The thoroughly mixed and ground material was heated at 920 °C for 24 hrs and 1200 °C for 36 hrs with intermediate grinding. Then the powders were pressed in form of rectangular pellets of dimension length=10 mm, width=5 mm and thickness=0.5mm and sintered at 1350 °C for 24 hrs. The structural and surface morphological characterization was done by powder X-ray diffraction (XRD) and scanning electron microscopy (SEM). The XRD data was refined by the Reitveld method using Full-Prof program. The magnetic phase characterization was carried out by measuring temperature and frequency dependent AC susceptibility of all the samples. The resistivity was measured in four contact configurations in the temperature range 4.2-350 K and the magnetoresistance in applied magnetic field up to 10 kOe was measured in the range 77-350 K. The magnetoresistance has been defined as MR = $(\rho_0 - \rho_H)/\rho_0$, $\rho_0$ and $\rho_H$ are resistivity measured at zero and magnetic field H respectively.

**Results and Discussion**

Since Gd is smaller than Nd, the average A-site cationic radius $<r_A>$ decreases, while $\sigma^2$ increases. The relationship between Gd concentration x, $<r_A>$ and $\sigma^2$ are linear. Structural characterization by XRD reveals that all the samples possess single phase orthorhombic structure and have very good crystallinity. Refinement of the XRD data carried out by the Reitveld method reveals that all samples have orthorhombic structure with space group P*bnm* and the lattice parameters are a > b > c (O' type orthorhombic). With the limit of the instrumental error the 'a' parameter is observed to decrease slightly with increasing Gd concentration (x), while 'b' and 'c' remain nearly constant. From the variation trend of the lattice parameters shows a gradual decrease in orthorhombicity with increasing Gd concentration. This could be understood in terms of intrinsic pressure in the orthorhombic lattice due to the substitution of smaller $Gd^{3+}$ (r=1.21 Å) cations for the larger $Nd^{3+}$ (r=1.27 Å) cations in the perovskite lattice. However, the variation in only one lattice parameter reflects the unidirectional nature of the pressure, which could be related to some specific/selective ordering of the $Gd^{3+}$ cations in the parent $Nd_{0.58}Sr_{0.42}MnO_3$ unit cell. SEM results show that the surface of these samples generally consists of spherical grains and well defined grain boundaries. Each grain is found to consist of several layers. The average grain size is found to be ~1 micrometer.

The PM-FM transition temperature ($T_C$) is observed to decrease with increasing $x/\sigma^2$ (see table 1). In the pristine sample (x=0), $\sigma^2$=0.00704 Å$^2$ and PM-FM transition occurs at $T_C$=275 K, which was found to gradually decrease as a function of increasing $\sigma^2$ to ≈82 K for x=0.42. The representative χ-T curves corresponding to x=0.00, 0.04, 0.12, 0.25 and 0.35 along the $\sigma^2$ values are plotted in figure 1 and the dependence of $T_C$ on $\sigma^2$ is shown in figure 3. In addition to this appreciable broadening of the PM-FM transition was observed at higher $\sigma^2$ and x. Thus Gd substitution at the Nd site has the twin effects on the PM-FM phase transition. The $T_C$ suppression and concomitant increase in the transition width with increasing $x/\sigma^2$ is explained in terms of the structural and microstructural changes. The substitution of smaller $Gd^{3+}$ for larger $Nd^{3+}$ cations causes a decrease in the average A-site cationic radius <$r_A$> and in addition it also increases $\sigma^2$ (the intrinsic size disorder). The decrease in <$r_A$> coupled with the enhanced size disorder ($\sigma^2$) causes a local deformation of the $MnO_6$ octahedra, which strengthens the Jahn-Teller (J-T) distortion that increases carrier localization due to reduced bandwidth.[6-8] Simultaneously, the smaller values of <$r_A$> results in a decrease in the Mn-O-Mn bond angles, which weakens the FM-DE of the $e_g$-electrons. The usual broadening of the FM-transition with increasing x is due to the successive lowering of the Mn-O-Mn bond angles. Thus decrease in <$r_A$> (increase in $\sigma^2$) is expected to enhance the competition between the two coexisting phenomena, viz., the JT distortion and the FM-DE[6-8] and hence may cause strong phase fluctuation at or in the vicinity of a critical value of <$r_A$>/$\sigma^2$.[2] Since carrier localization is explicitly reflected in electrical transport measurement, the nature of the temperature dependent resistivity is expected to change in the vicinity of the critical <$r_A$>/$\sigma^2$. At intermediate Gd contents (x=0.20-0.30) the frequency dependence of AC susceptibility shows evidence of a glassy behavior.

The temperature dependence of resistivity, along with the $\sigma^2$ values is shown in figure 3. The insulator-metal transition (IMT) of the Gd free sample is $T_{IM} \approx 223$ K and like the $T_C$ it is also gradually observed to decrease with increase in $<r_A>/\sigma^2$. The $T_{IM}$ of x=0.35 sample was ~83 K and the x=0.42 sample did not show IMT at all. The variation of IMT as a function of $\sigma^2$ is shown in figure 3 and also listed in table 1. However, unlike the PM-FM transition which broadened with increasing $<r_A>/\sigma^2$, the IMT was observed sharpen. The sharpening of the IMT is clearly visible in the ρ-T curves presented in figure 2.

It is interesting to note that both $T_C$ and $T_{IM}$ decrease gradually as $\sigma^2$ (or x) increases. In the pristine sample ($\sigma^2$=0.00704 Å$^2$), the difference between $T_C$ and $T_{IM}$ is large ~50 K ($T_C>T_{IM}$). At $\sigma^2$=0.00933 Å$^2$ (x=0.20), both $T_C$ and $T_{IM}$ attain the same value and at further higher variances the difference between them again increases slightly ($T_{IM}<T_C$). In polycrystalline manganites, the difference between the $T_C$ and $T_{IM}$ is generally attributed to the grain boundary (GB) disorder.[14-16] Therefore, at lower values of $\sigma^2$ the difference between $T_C$ and $T_{IM}$ is due to the GB disorder. However, near equality of $T_C$ and $T_{IM}$ at intermediate values of $\sigma^2$ could be understood in terms of the quenched disorder due to the size mismatch. As the size disorder increases the quenched disorder induces phase fluctuation and enhanced phase competition. The competing phases in the presented case are expected to be FM-metal and AFM-CO-insulator.[2] Consequently, the competition between these two disorders, viz., the GB disorder and the quenched disorder is responsible for the observed variation in $T_C$ and $T_{IM}$. At lower values of x, e.g., the Gd free sample the large difference in $T_C$ and $T_{IM}$ shows that the GB disorder dominates over the size mismatch induced quenched disorder measured by $\sigma^2$. As the Gd concentration increases the GB disorder remains unaffected while the value of $\sigma^2$ increases and enhances the phase fluctuation. Since $Nd_{0.58}Sr_{0.42}MnO_3$ has a FM-M ground state while $Gd_{0.58}Sr_{0.42}MnO_3$ is an AFM-CO insulator, the partial substitution of Nd by Gd could result in creation of AFM-I/CO-I phase in the parent FM-M matrix and hence enhance phase fluctuation. These phase fluctuations are expected to increase with increasing Gd content and maximize around intermediate values of x, e.g., x=0.20-0.30. So at these Gd concentrations the phase fluctuations is expected to be the maximum and dominate over the GB disorder resulting in near equality of $T_C$ and $T_{IM}$. Near the highest value of $\sigma^2$ the $T_{IM}$ becomes smaller than the corresponding $T_C$. This could be due to the appearance of some AFM-CO-OO cluster in the FM matrix.

The temperature and magnetic field dependent magnetoresistance (MR) of all samples was measured. The representative plot showing the variation of MR (H=3 kOe) as a function of temperature is shown in figure 4. It is well known that the MR in polycrystalline samples has two contributions: (i) the intrinsic component that arises due to the FM-DE and peaks around $T_C/T_{IM}$ and (ii) the extrinsic part that is determined by the grain boundaries and increases monotonously as the temperature is lowered.[14-16] As seen in the figure 4, with increasing x and hence $\sigma^2$, the MR is observed to increase. The intrinsic MR is observed to increase from ~3 % to 35% as the variance increases from 0.00704 to 0.009857 Å$^2$ and then decreases for larger values of $\sigma^2$. In contrast, the extrinsic GB part remains nearly constant and then increase slightly, thus it has a weak dependence on the $\sigma^2$. The variation of the peak MR measured at 3 kOe and 10 kOe and MR at T=77 K is shown in the inset of figure 4. The intrinsic MR that shows a maximum around $T_C/T_{IM}$ is expected to be the highest in the vicinity of the maximum phase fluctuation. To verify this we measured the temperature dependent MR at different magnetic fields for all the samples. The variation of peak MR (maximum MR around $T_C/T_{IM}$) with $\sigma^2$ shows that maximum low field MR≈35 % at H=3 kOe (68 % at H=10 kOe) is centered around $\sigma^2 = 0.009857$ Å$^2$ (x=0.25). This shows that huge intrinsic MR can be obtained at relatively higher temperatures and lower magnetic fields in the region of enhanced phase fluctuations.

**Conclusions**

We have studied the effect of the quenched disorder arising from the size disorder (measured by the variance $\sigma^2$) at the RE site in polycrystalline $Nd_{0.58-x}Gd_xSr_{0.42}MnO_3$. Although both $T_C$ and $T_{IM}$ are observed to decrease with increasing $\sigma^2$ the former becomes broad while the later is sharpened. This feature is attributed to phase fluctuation due to the size disorder caused by Gd doping. The large difference in the $T_C$ and $T_{IM}$ ($T_C<T_{IM}$) at lower values of $\sigma^2$ suggests the dominance of the grain boundary disorder over the quenched disorder/phase fluctuation. Gradual sharpening of the IMT with increasing $\sigma^2$ followed by very sharp metallic transition around $\sigma^2 = 0.009857$ Å$^2$ shows that dominance of quenched disorder/phase fluctuation. At further higher $\sigma^2$ the $T_{IM}$ again becomes smaller than $T_C$ and IMT finally vanishes. Large low field MR observed in the vicinity of $\sigma^2 = 0.009857$ Å$^2$ demonstrates the importance of quenched disorder/phase fluctuation in manganites.

**Acknowledgements**

MKS and RP thankfully acknowledge CSIR and UGC-New Delhi for award of research fellowship.

**References**

1. C. N. R. Rao, and B. Raveau, (eds.) Colossal Magnetoresistance, Charge Ordering and Related Properties of Manganese Oxides (Singapore: World Scientific) (1998).
2. Y. Tokura, Rep. Prog. Phys. **69**, 797 (2006).
3. E. Dagotto, T. Hotta, and A. Moreo, Phys. Rep. **344**, (2001).
4. E. Dagotto, New Journal of Physics **7,** 67 (2005).
5. A. Moreo, S. Yunoki, and E. Dagotto, Science **283**, 2034 (1999).
6. L. M. Rodriguez-Martinez, and J. P. Attfield, Phys. Rev. B **54**, 15622 (1998).
7. J. P. Attfield, Chem. Mater. **10**, 3239 (1998).
8. L. M. Rodriguez-Martinez, and J P Attfield, *Phys. Rev.* B **63**, 024424 (2000).
9. R. D. Shannon, Acta Crystallogr., Sect. A: Cryst. Phys., Diffr., Theor. Gen. Crystallogr. **32**, 751(1976).
10. K. R. Mavani, and P. L. Paulose, Appl. Phys. Lett. **86**, 162504 (2005).
11. K. F. Wang, Y. Wang, L. F. Wang, S. Dong, H. Yu, Q. C. Li, J.-M. Liu, and Z. F. Ren, Appl. Phys. Lett. **88**, 152505 (2006)
12. K. F. Wang, F. Yuan, and S. Dong, D. Li, Z. D. Zhang, Z. F. Ren, and J.-M. Liu, Appl. Phys. Lett. **89**, 222505 (2006).
13. N. Rama and M. S. Ramachandra Rao, V. Sankaranarayanan, P. Majewski, S. Gepraegs, M. Opel, and R. Gross, Phys. Rev B **70**, 224424 (2004).
14. A. Gupta, G. Q. Gong, G. Xiao, P. R. Dumcombe, P. Lecoeur, P. Trouilloud, Y. Y. Wang, V. P. Dravid, and J. Z. Sun, Phys. Rev. B **54,** R15629 (1996).
15. X. L. Wang, S. X. Dou, H. K. Liu, M. Ionescu, and B. Zeimetz, Appl. Phys. Lett. **73**, 396 (1998).
16. P. K. Siwach, H. K. Singh, and O. N. Srivastava, J. Phys. Cond. Matter **20**, 273201 (2008).

**Figure Captions**

Figure 1   Variation of AC susceptibility ($\chi$) as a function of Temperature for $Nd_{0.58-x}Gd_xSr_{0.42}MnO_3$.

Figure 2   Variation of resistivity as a function of Temperature of $Nd_{0.58-x}Gd_xSr_{0.42}MnO_3$. The corresponding variances $\sigma^2$ (in Å$^2$) have been indicated.

Figure 3   Variation of $T_C$ and $T_{IM}$ with variance $\sigma^2$ (with unit Å$^2$).

Figure 4   Temperature dependence of MR measured at magnetic field H=3 kOe. The inset shows the variation of peak MR (%) with $\sigma^2$ (with unit Å$^2$).

**Table 1**

| Gd Content (x) | $<r_A>$ (Å) | $\sigma^2$ (x10$^{-3}$Å$^2$) | $T_C$ (K) | $T_{IM}$ (K) |
|---|---|---|---|---|
| 0 | 1.3414 | 7.04 | 275 | 223 |
| 0.04 | 1.339 | 7.521 | 263 | 213 |
| 0.08 | 1.3366 | 7.99 | 248 | 213 |
| 0.12 | 1.3342 | 8.448 | 223 | 203 |
| 0.16 | 1.3318 | 8.895 | 198 | 188 |
| 0.20 | 1.3294 | 9.33 | 168 | 168 |
| 0.25 | 1.3264 | 9.857 | 133 | 118 |
| 0.30 | 1.3234 | 10.366 | 117 | 105 |
| 0.35 | 1.3204 | 10.858 | 108 | 83 |
| 0.42 | 1.3162 | 11.516 | 83 | - |

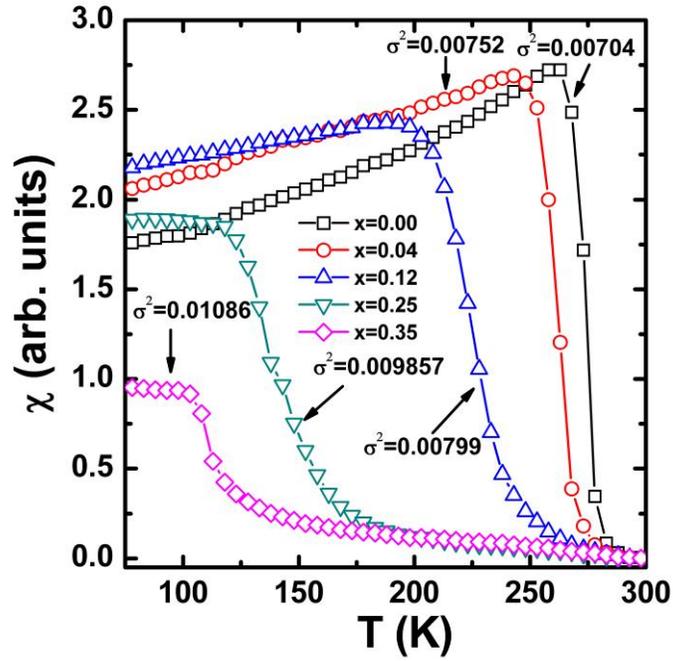

Fig. 1

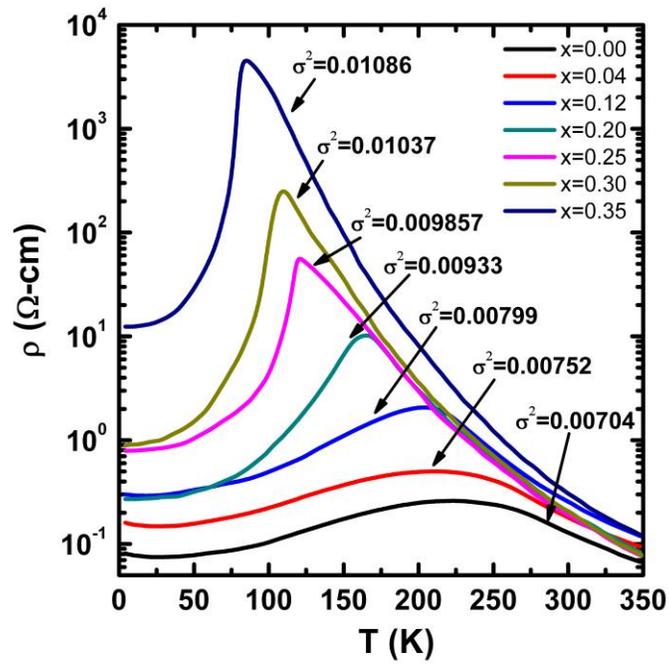

Fig. 2

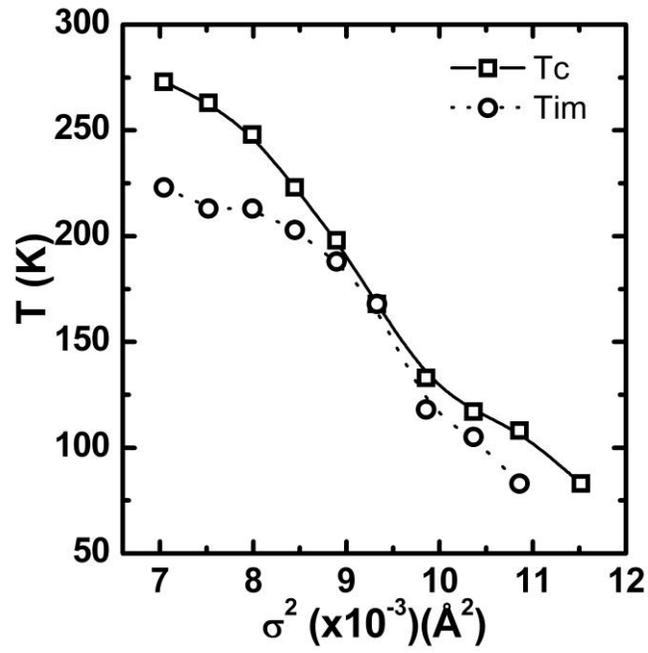

Fig. 3

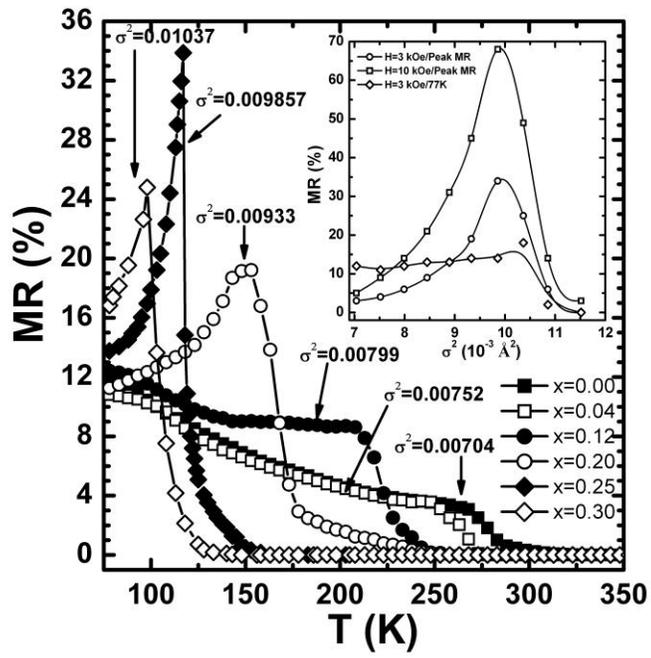

Fig. 4